\documentclass[slac_one]{revtex4}
\usepackage{graphicx}
\usepackage{fancyhdr}
\usepackage{units}
\usepackage[percent]{overpic}
\newcommand{\bfig}[1][t!]{\begin{figure}[#1] \begin{center}}
\newcommand{\efig}{\end{center} \end{figure}}
\newcommand{\twoPlots}[3][t!]{\bfig[#1] \begin{minipage}[t]{0.48\linewidth} \vspace{0pt} \centering #2 \end{minipage} \hfill \begin{minipage}[t]{0.48\linewidth} \vspace{0pt} \centering #3 \end{minipage} \efig}
\newcommand{\cmsofT}[1]{$\sqrt{s} = \unit[#1]{TeV}$}

\newcommand{\etain}[1]{$|\eta|$~$<$~$#1$}
\newcommand{\cms}{\sqrt{s}}
\newcommand{\dndeta}{dN_{\rm ch}/d\eta}
\newcommand{\pbar}{\bar{\mbox{p}}}
\newcommand{\pt}{p_{\rm T}}

\begin{document}

\title{Minimum-Bias and Early QCD Physics in ALICE} 

%

\author{J. F. Grosse-Oetringhaus for the ALICE collaboration}
\email{Jan.Fiete.Grosse-Oetringhaus@cern.ch}
\affiliation{CERN, Geneva, Switzerland}
%

\maketitle

\thispagestyle{fancy}


\section{Introduction}

A Large Ion Collider Experiment (ALICE) \cite{alice} is the dedicated heavy-ion experiment at the Large Hadron Collider (LHC). In addition to its heavy-ion physics program, it also has a rich proton--proton physics program benefiting from a detector with a low momentum cut-off ($\pt \sim \unit[50]{MeV/\emph{c}}$) and a small material budget (about 11\% of a radiation length until the outer wall of the main tracking detector, the Time-Projection Chamber). ALICE has excellent means of particle identification (PID) with methods ranging from specific energy loss and time of flight to transition and Cherenkov radiation. The good primary and secondary vertex resolution allows for measurements of strangeness and heavy flavor with low backgrounds.

ALICE has taken proton--proton collision data at 0.9, 2.36, and \unit[7]{TeV}. In this article results of the first minimum-bias and soft-QCD measurements are presented.

\section{Charged-Particle Multiplicity}

\twoPlots[b!]{
\includegraphics[width=\textwidth]{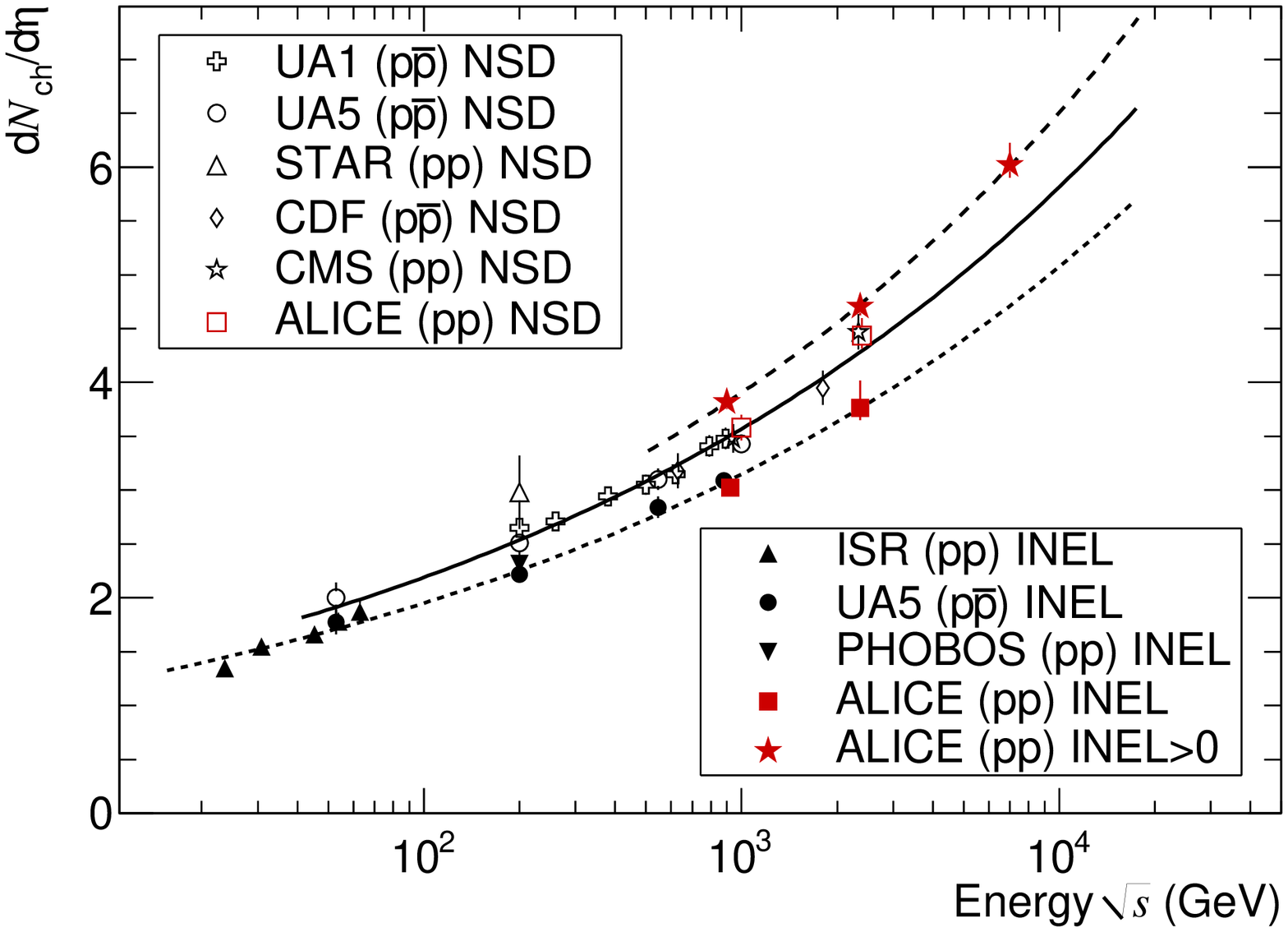}
\caption{Charged-particle pseudorapidity density in the central pseudorapidity region \etain{0.5}
 for INEL and NSD collisions, and in \etain{1} for inelastic collisions with at least one charged particle in that region (INEL$>$0), as a function of the center-of-mass energy. The lines indicate a
fit using a power-law dependence on collision energy. Note that data points at the same energy have been slightly shifted horizontally for visibility. The figure is from \cite{alice7}.}
\label{dndeta_vs_sqrt}
}{
\includegraphics[width=\textwidth]{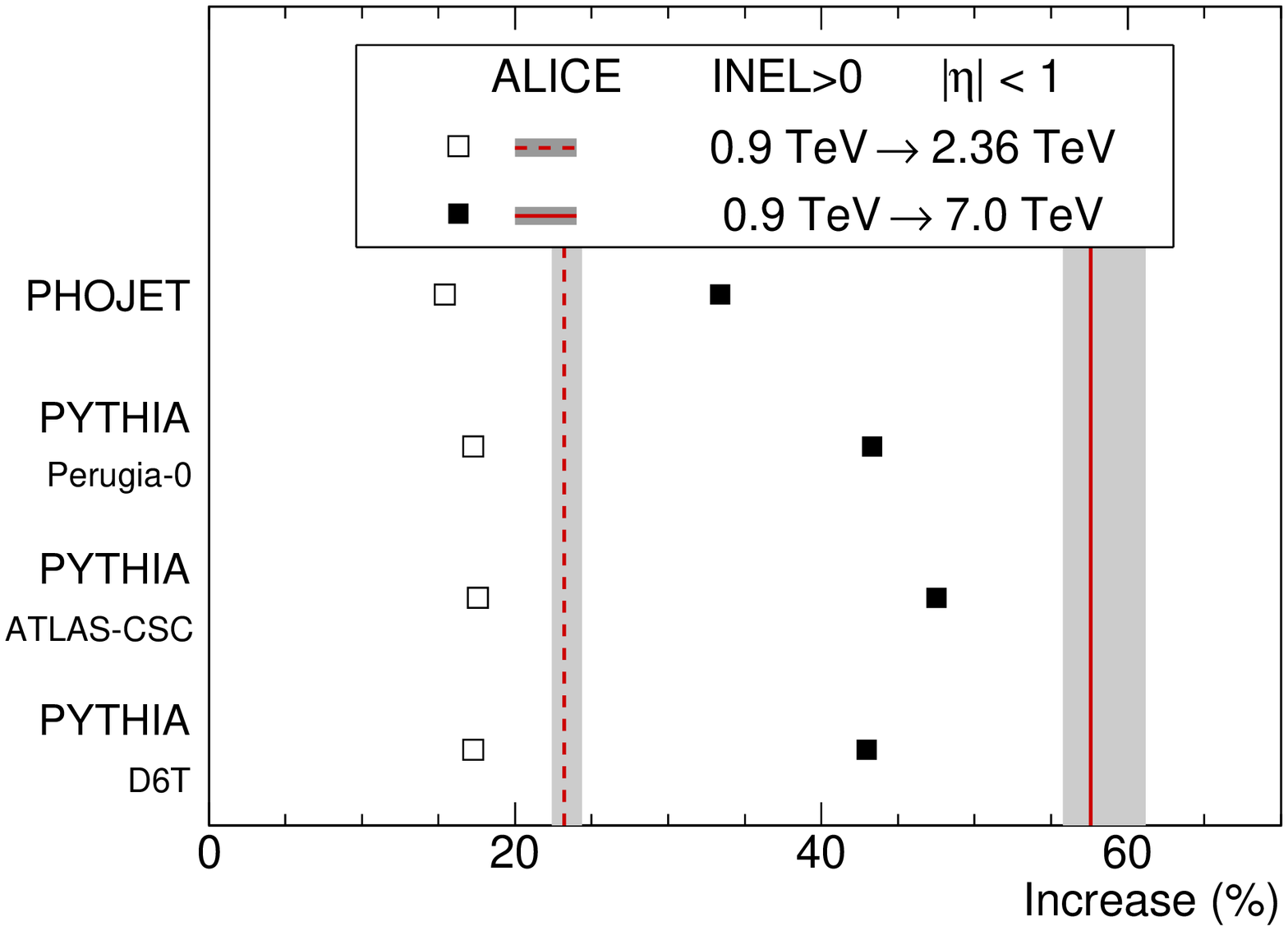}
\caption{Relative increase of the charged-particle pseudorapidity density, for inelastic collisions having at least one charged particle in \etain{1}, between $\cms =0.9$~TeV and 2.36~TeV (open squares) and between $\cms = 0.9$~TeV and 7~TeV (full squares), for various models. Corresponding ALICE measurements are shown with vertical dashed and solid lines, respectively.
The figure is from \cite{alice7}.} \label{dndeta_growth}
}

ALICE has measured the charged-particle multiplicity for inelastic (INEL) and non-single-diffractive (NSD) collisions \cite{alice1,alicemult} as well as for events that have at least one charged particle in \etain{1} (INEL$>$0) \cite{alice7}.
The minimum-bias trigger used in this analysis accepts 95--97\% of all inelastic events which was estimated with MC generators~\cite{pythia,phojet} and the detector simulation and analysis framework AliRoot~\cite{aliroot}.

Figure~\ref{dndeta_vs_sqrt} presents the average $\dndeta$ at mid-rapidity for INEL, NSD, and INEL$>$0 events measured by ALICE as well as data from other experiments.
Figure~\ref{dndeta_growth} shows the relative increase in the average multiplicity from \cmsofT{0.9} to 2.36 and \unit[7]{TeV} compared to commonly used MC generators. Predictions are shown from PYTHIA with the tunes D6T \cite{d6t}, ATLAS-CSC \cite{atlascsc} and Perugia-0 \cite{perugia0}. The measured increase is significantly larger than predicted: for the increase from 0.9 to \unit[7]{TeV} we find a value of $57.6\,\% \pm 0.4\,\%(\emph{stat.}) ^{+3.6}_{-1.8}\,\%(\emph{syst.})$ compared to predictions in the range of 33\% to 48\%. The prediction closest to the measured value is PYTHIA with the tune ATLAS-CSC.

\bfig
    \includegraphics[width=0.48\textwidth]{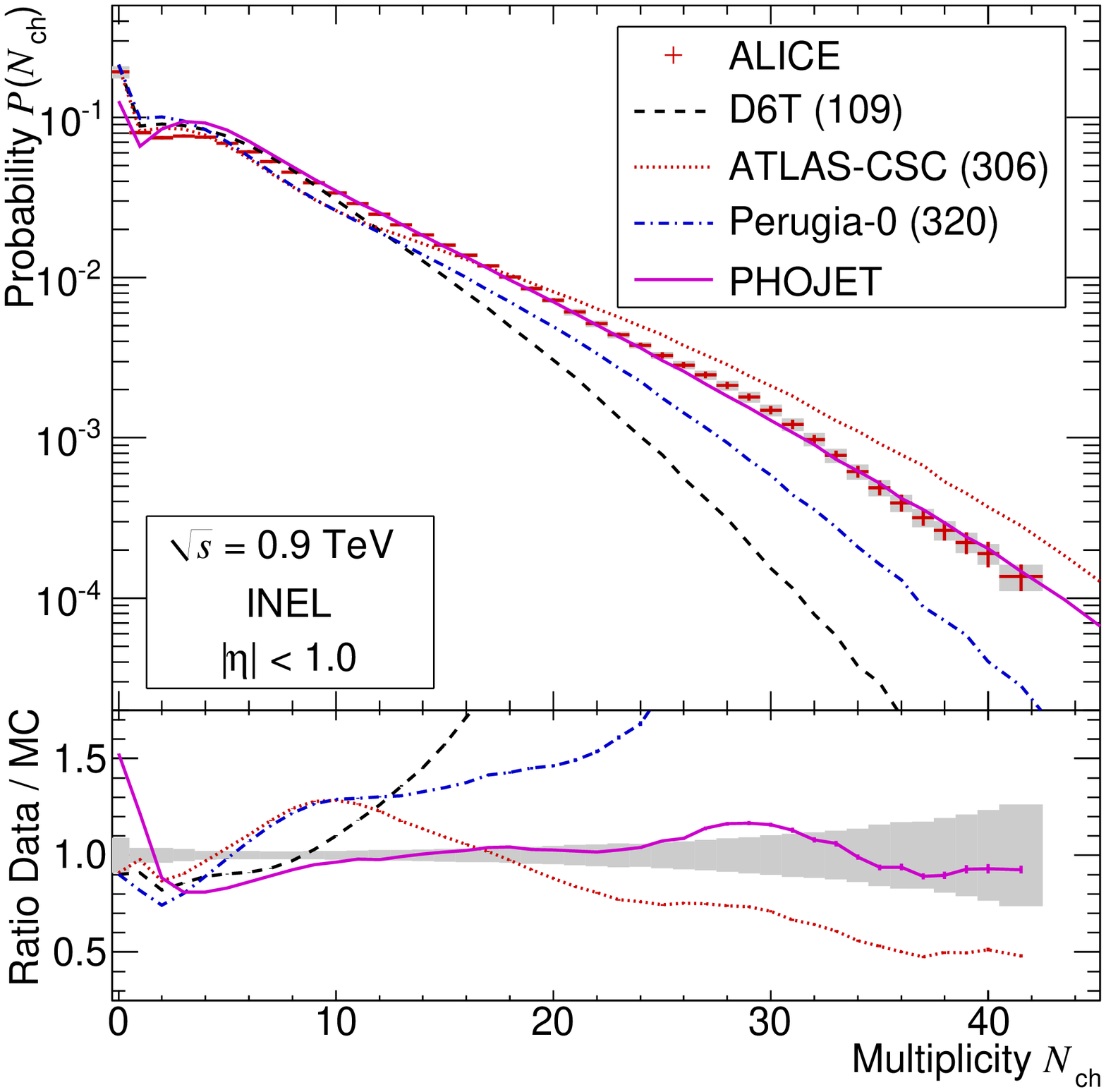}
    \hfill
    \includegraphics[width=0.48\textwidth]{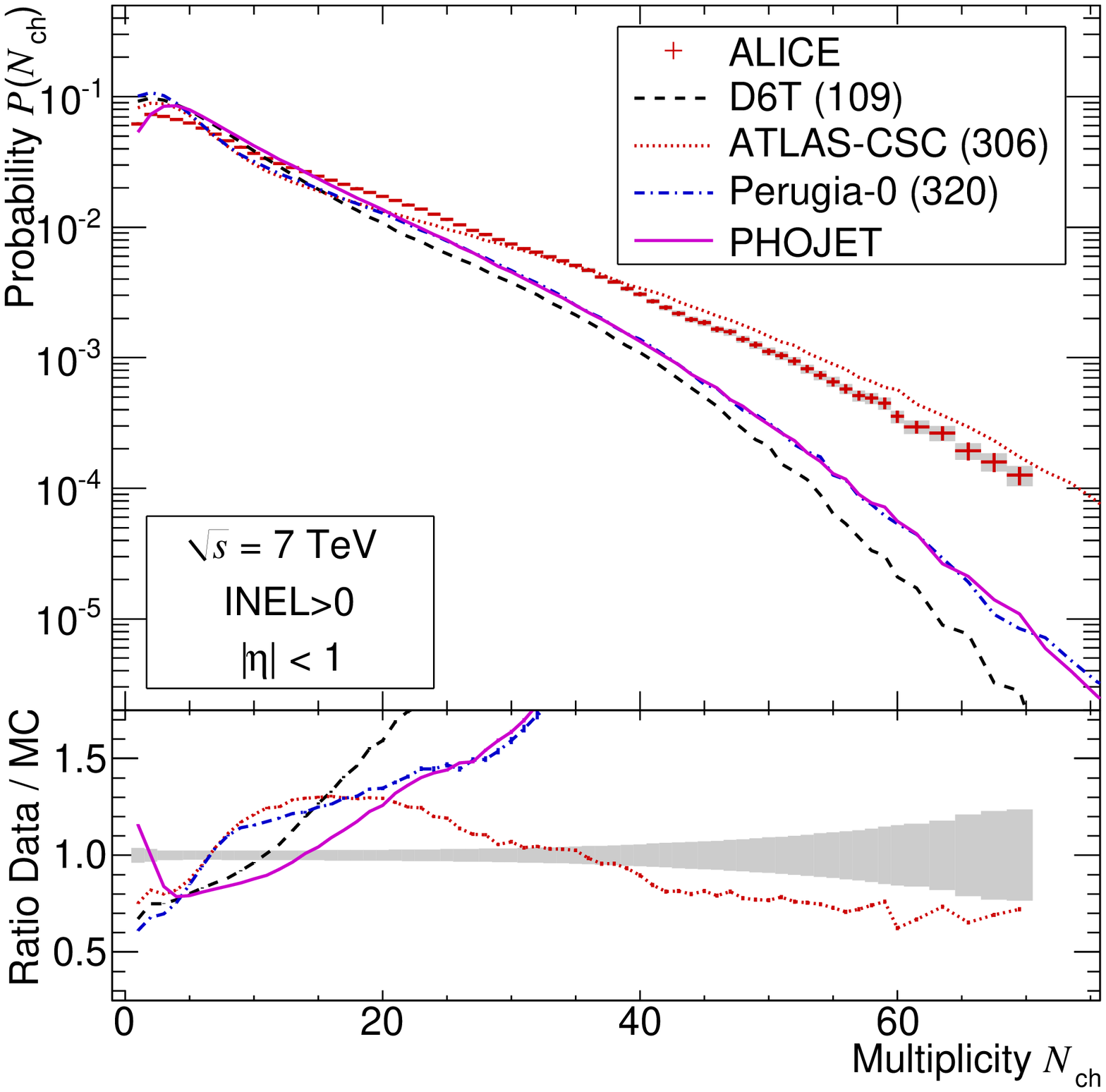}
    \caption{The figure shows the multiplicity distribution of charged particles in \etain{1} at \unit[0.9]{TeV} (left panel) and at \unit[7]{TeV} (right panel) compared to MC generator predictions. The lower panels show the ratio data over MC. The figures are from \cite{alicemult,alice7}. \label{fig_mult}}
\efig

Figure~\ref{fig_mult} presents the charged-particle multiplicity distribution in \etain{1} at 0.9 and \unit[7]{TeV} for inelastic collisions. These are compared to the model predictions mentioned earlier. At the lower energy, PHOJET describes the spectrum best, at the higher energy, only ATLAS-CSC is close to the data (with deviations in the multiplicity region of 10--25). The three other tunes differ significantly at \unit[7]{TeV}; in particular the tail of the measured charged-particle distribution is much wider.

\section{Transverse-momentum distributions}

\twoPlots{
\includegraphics[width=0.85\textwidth]{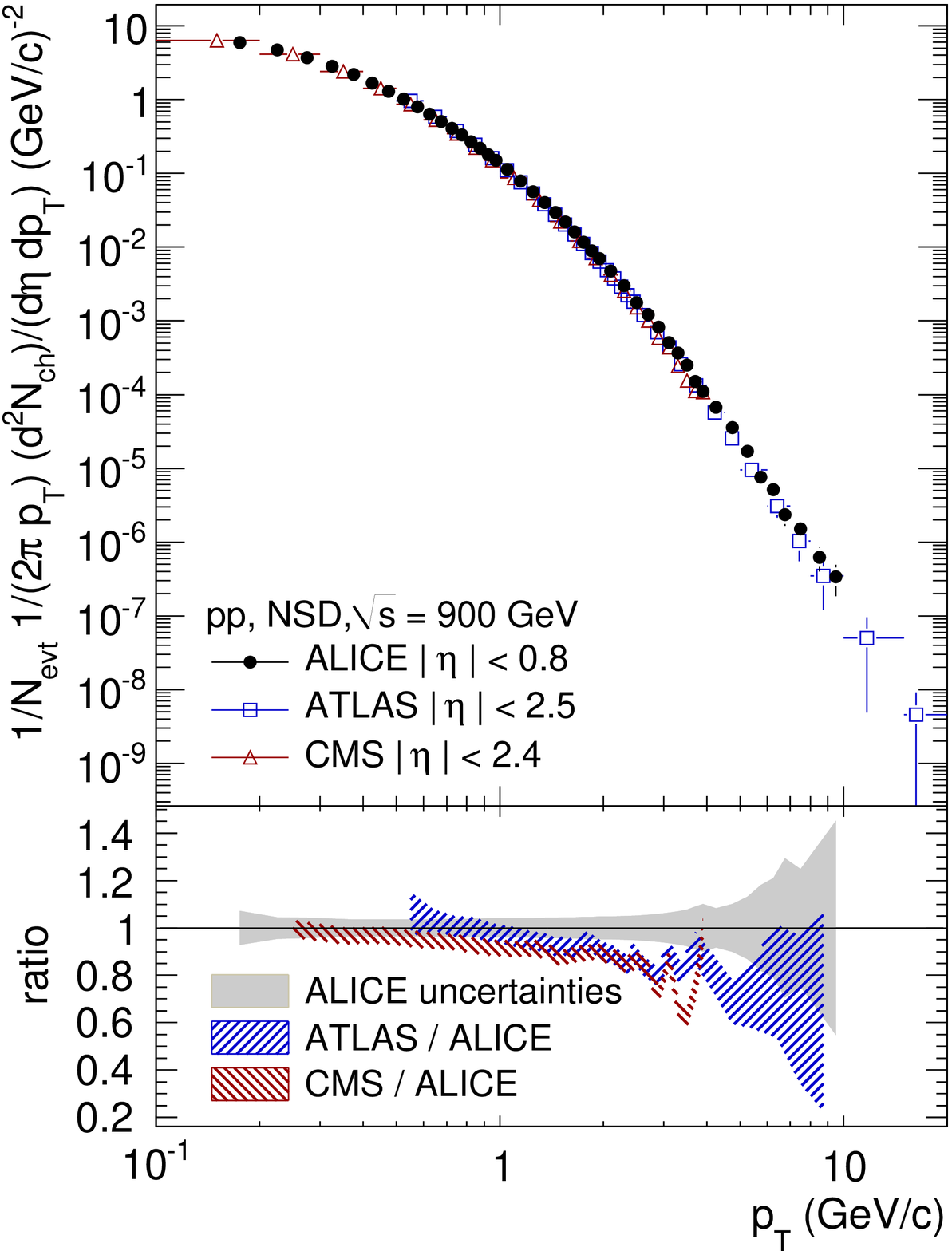}
\caption{Normalized differential primary charged-particle yield in NSD pp collisions at \unit[0.9]{TeV}, averaged in \etain{0.8}. The ALICE data are compared to results from ATLAS and CMS in pp at the same energy \cite{altaspt,cmspt}. The lower panel shows the ratio. Note that the measurements are in different $\eta$-intervals. The figure is from \cite{alicept}.}
\label{ptspectrum}
}{
\includegraphics[width=0.85\textwidth]{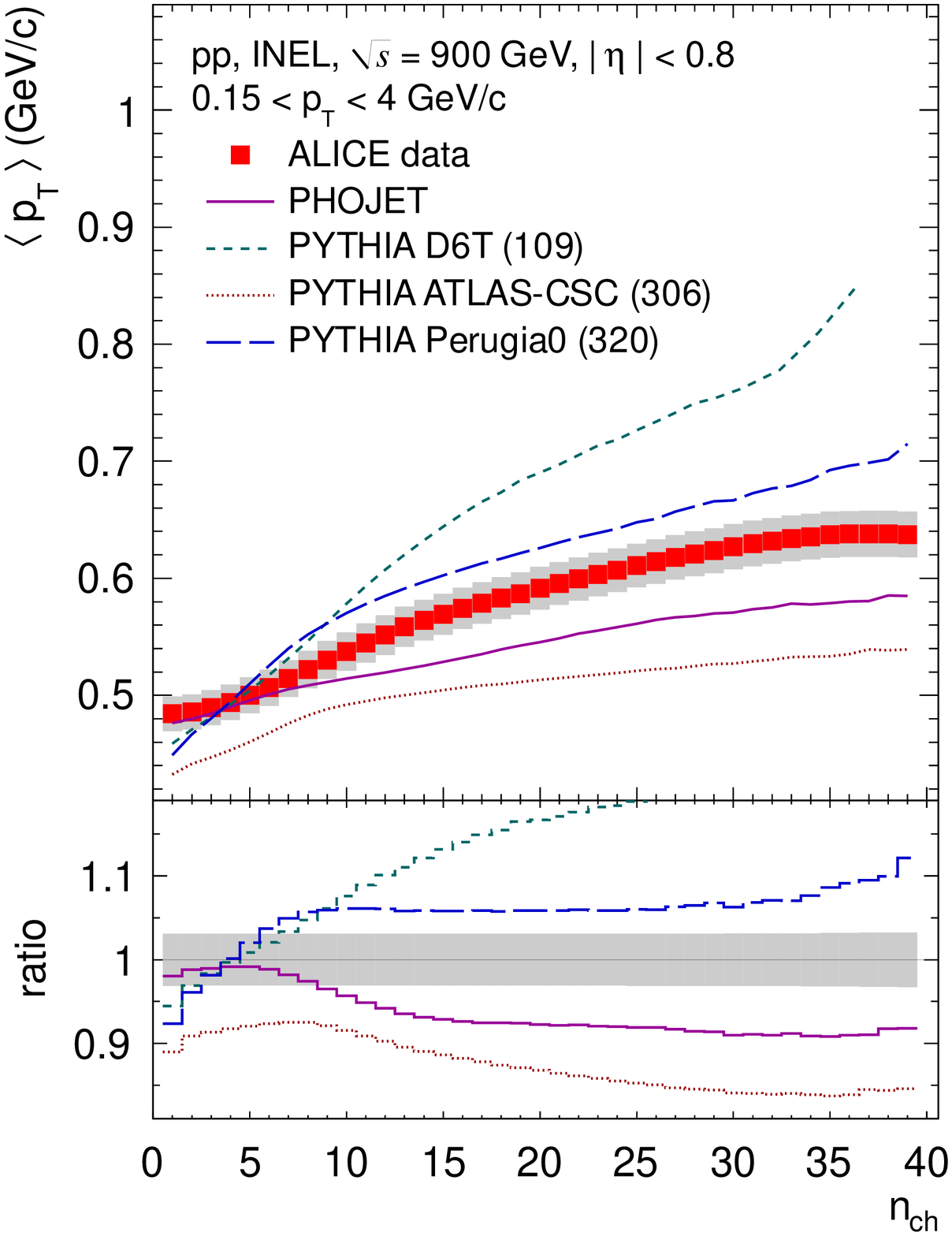}
\caption{The average transverse momentum of charged particles within $0.15 < \pt < \unit[4]{GeV/\emph{c}}$
of INEL pp events at \unit[0.9]{TeV} as a function of the charged-particle multiplicity is presented.
In the lower panel, the ratio MC over data is shown.
The figure is from \cite{alicept}.} \label{ptvsmult}
}

The transverse-momentum distribution $dN_{\rm ch}/d\pt$ of INEL and NSD events at \unit[0.9]{TeV} was measured, together with the correlation of the average-$\pt$ with the multiplicity in the event \cite{alicept}. Figure~\ref{ptspectrum} shows the comparison of the ALICE measurement for NSD events to similar results from ATLAS and CMS. While the ALICE result is in the pseudorapidity range \etain{0.8}, ATLAS and CMS measure in a larger acceptance window. One can see that the spectrum gets harder towards mid-rapidity or towards smaller rapidity windows.

Figure~\ref{ptvsmult} presents the average-$\pt$ for particles within $0.15 < \pt < \unit[4]{GeV/\emph{c}}$ as a function of the event multiplicity, compared to MC predictions. PYTHIA tune Perugia-0 is closest to the data but has some clear deviations. It should be noted that Perugia-0 describes the same distribution well for particles within $0.5 < \pt < \unit[4]{GeV/\emph{c}}$ (plot not shown) indicating the difficulty to model the low-momentum region and thus the importance to measure it well.

\section{Two-Pion Bose-Einstein correlations}

\bfig
    \hspace{0.02\textwidth}
    \includegraphics[width=0.46\textwidth]{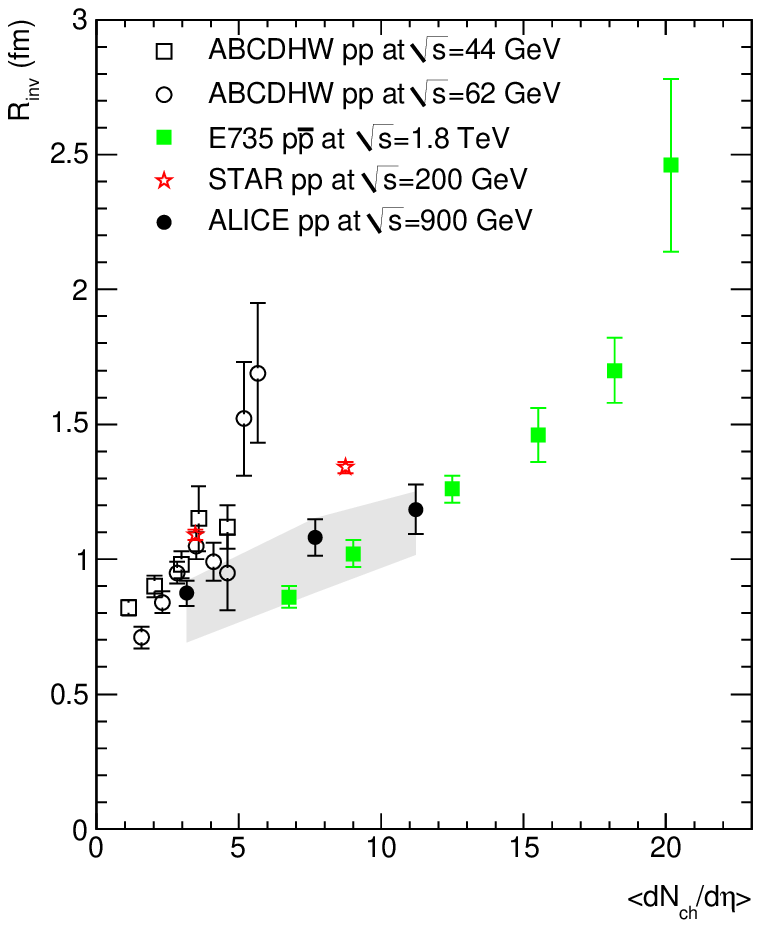}
    \hfill
    \includegraphics[width=0.44\textwidth]{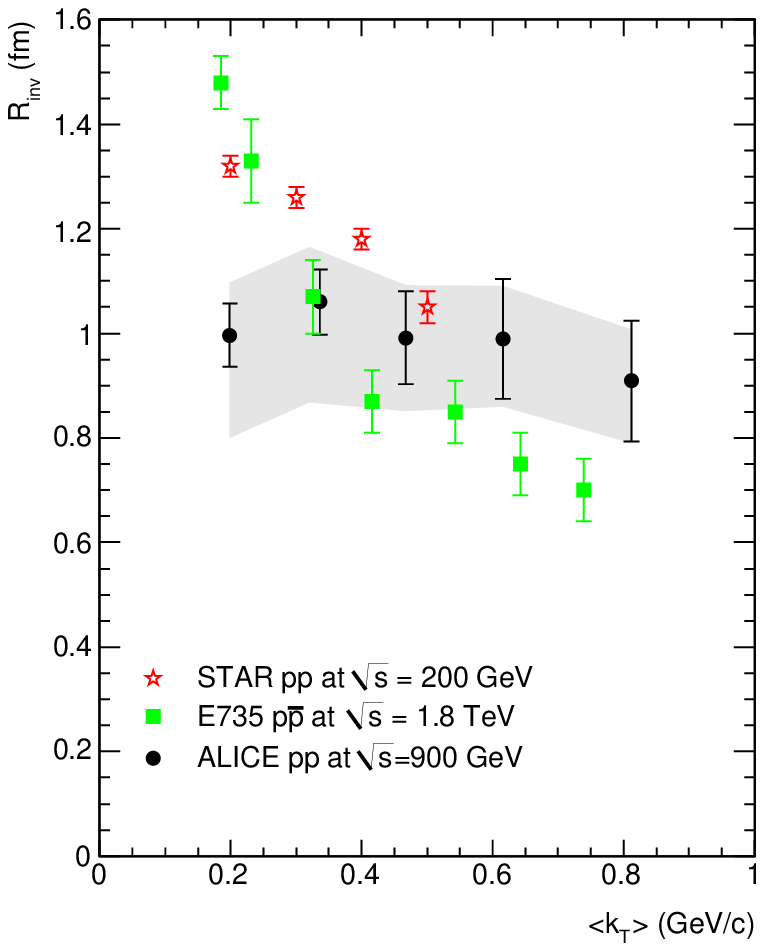}
    \hspace{0.02\textwidth}
    \caption{One-dimensional Gaussian HBT radius in pp collisions at \cmsofT{0.9} shown as a function of the charged-particle multiplicity at mid-rapidity (left panel) and the average transverse momentum $k_{\rm T}$ (right panel).
The ALICE measurement is compared to data taken at other experiments. The figures are from \cite{hbt}. \label{twopion}}
\efig

The measurement of Bose-Einstein enhanced correlations of identical particle pairs allows to assess the size of the emitting source. These measurements have become a precision tool in heavy-ion collisions to probe the geometry and dynamics of the emitting system~\cite{hbtHI}, but were initially carried out in smaller systems like p$\pbar$~\cite{hbtpp} and also $e^+e^-$~\cite{hbtee}. The two-particle correlation function $C(q)$ is measured, where $q = p_2 - p_1$ is the four-momentum difference between the two particles. ALICE has measured $C(q)$ as a function of multiplicity and average transverse momentum $k_{\rm T} = |p_{\rm T,1} + p_{\rm T,2}| / 2$ \cite{hbt}. Fitting the correlation function with a Gaussian allows to extract the source size $R_{\rm inv}$ which is presented in Figure~\ref{twopion} as a function of multiplicity and $k_{\rm T}$. The source size increases as a function of multiplicity, consistent with the measurement of other experiments. As a function of $k_{\rm T}$ the source size is rather flat, not consistent with the observations of STAR and E735 at different energies. It should be noted that the dependency as a function of $k_{\rm T}$ is very sensitive to the choice of baseline which corresponds to the correlation function in absence of Bose-Einstein enhancement. For details see \cite{hbt}.

\section{Mid-rapidity antiproton-to-proton ratio}

\bfig
    \includegraphics[width=0.5\textwidth]{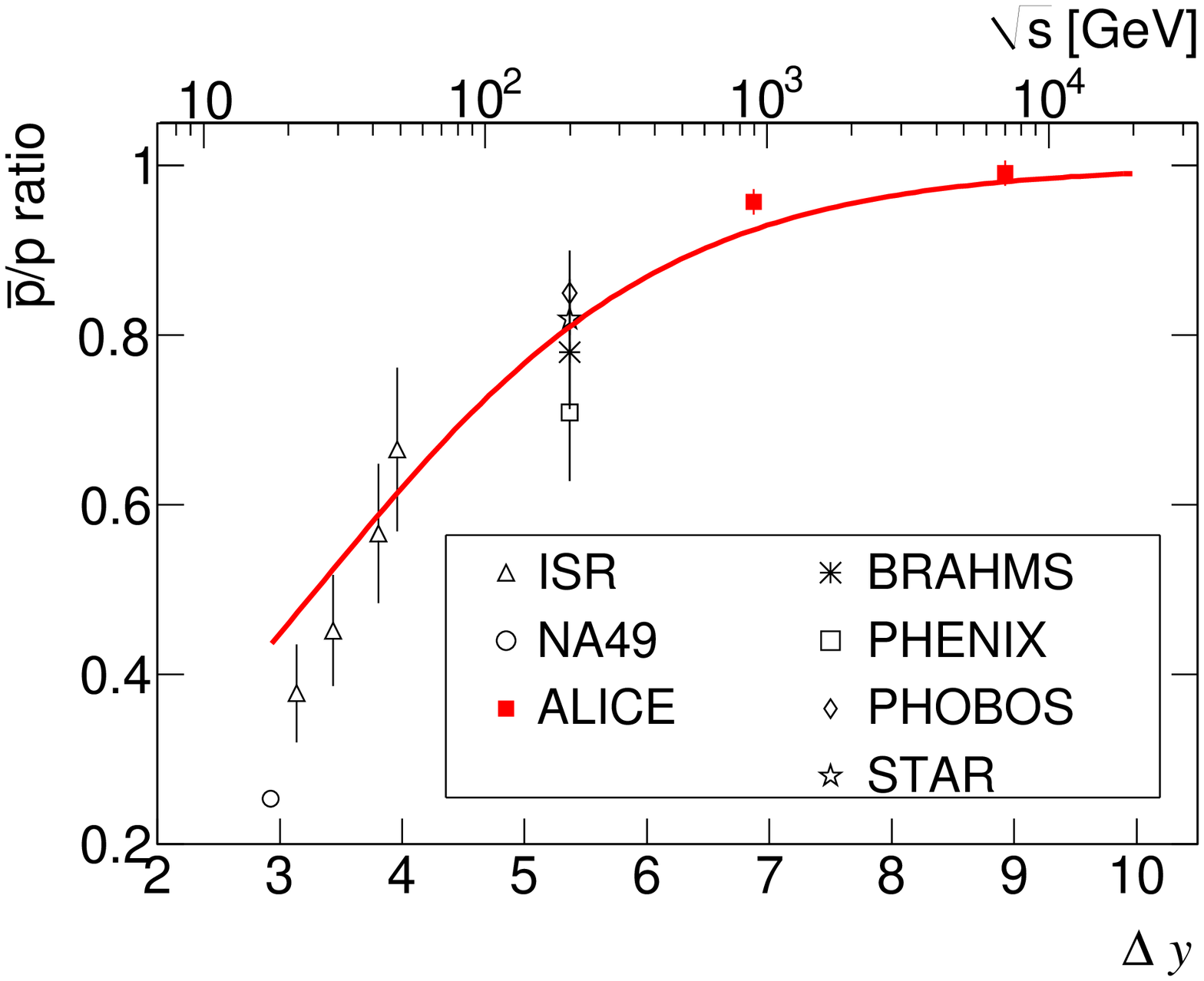}
    \caption{Central rapidity $\pbar$/p yield ratio as a function of the rapidity interval from the beam $\Delta y$ (lower axis) and center-of-mass energy $\cms$ (upper axis). Error bars correspond to the quadratic sum of statistical and systematic uncertainties for the RHIC and LHC measurements and to statistical errors otherwise. The figure is from \cite{aliceppbar}.} \label{ppbar_vs_sqrts}
\efig

The incoming state of a pp collision has a net baryon number of 2. It is interesting to study the redistribution of this baryon number in the final state and in particular how far towards mid-rapidity baryon number can be transferred. The baryon-number transfer can be modeled as the breaking of one or several strings between the valence quarks and the so-called string junction identified with the baryon number. A large rapidity loss $\Delta y= y_{\rm baryon} - y_{\rm beam}$ requires processes described by Regge trajectories with large intercepts, see e.g. \cite{ppbartransfer}.

ALICE has measured the antiproton over proton yield ratio at mid-rapidity \cite{aliceppbar}, whose largest systematic uncertainty is due to the different absorption cross section of $\pbar$ and p, requiring that the material budget of the detector is well understood.
The measured ratio was found to be independent of $\pt$ in the range of $0.5 < \pt < \unit[1]{GeV/\emph{c}}$. Comparison to MC simulations shows that predictions involving enhanced baryon-number transfer do not reproduce the data (plot not shown).
Figure~\ref{ppbar_vs_sqrts} shows the ratio as a function of $\cms$ or equivalently beam rapidity together with results from lower energies. The data are fitted with a parametrization describing p and $\pbar$ production, i.e., combining pair production and baryon-number transfer. The fit is successful with a junction intercept of 0.5, for details see \cite{aliceppbar}. In conclusion, there is little room left for production mechanisms which transport baryon number over a large rapidity interval.

\section{Identified Particle Spectra \& Strangeness}

\twoPlots{
\begin{overpic}[width=\textwidth,trim=0 0 1cm 0.8cm,clip=true]{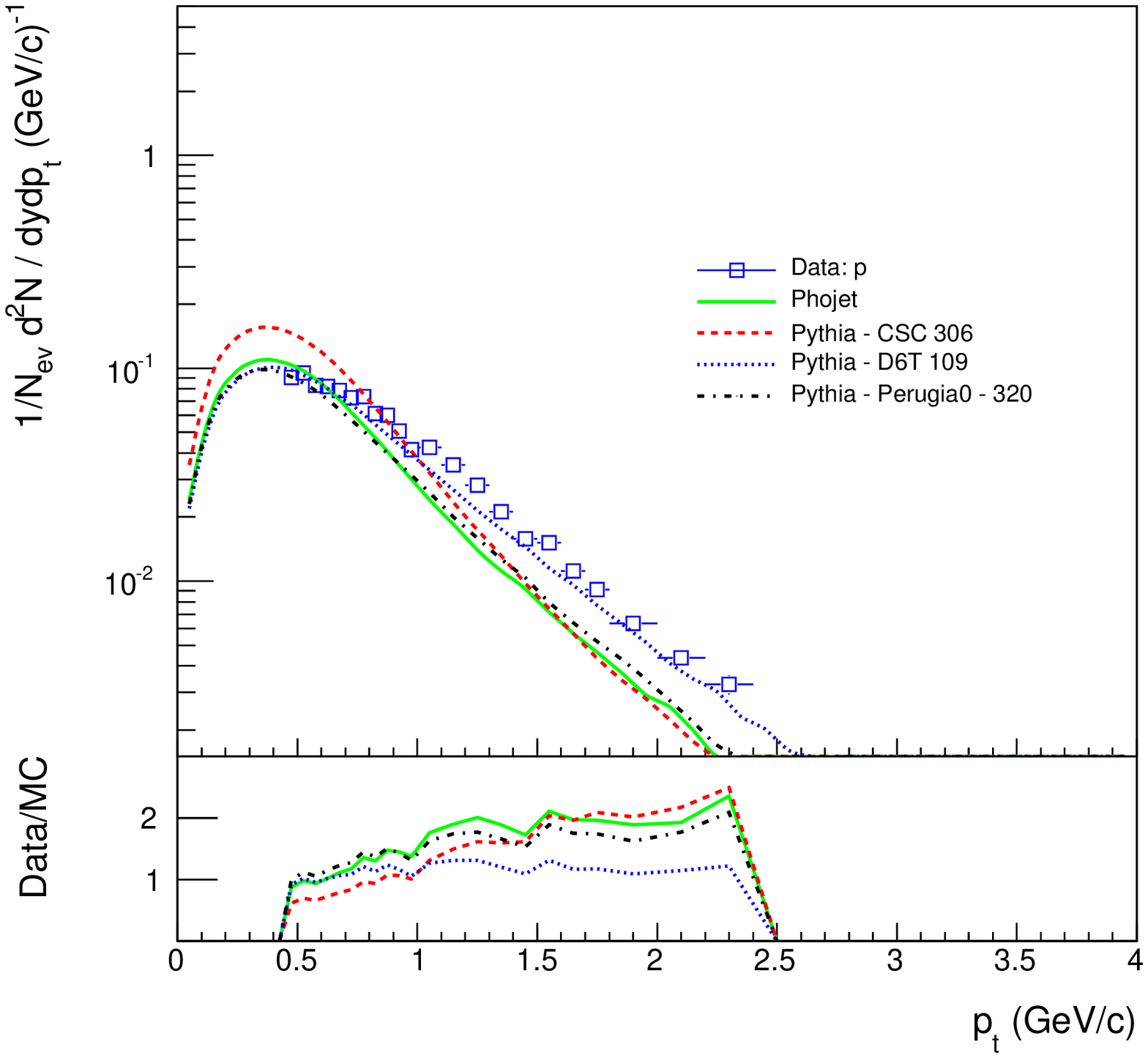}
\put(60,85){ALICE Preliminary}
\end{overpic}
\caption{Shown is the yield of protons as a function of $\pt$ compared to MC predictions. The bottom panel shows the ratio data over MC.}
\label{protons}
}{
\begin{overpic}[width=\textwidth]{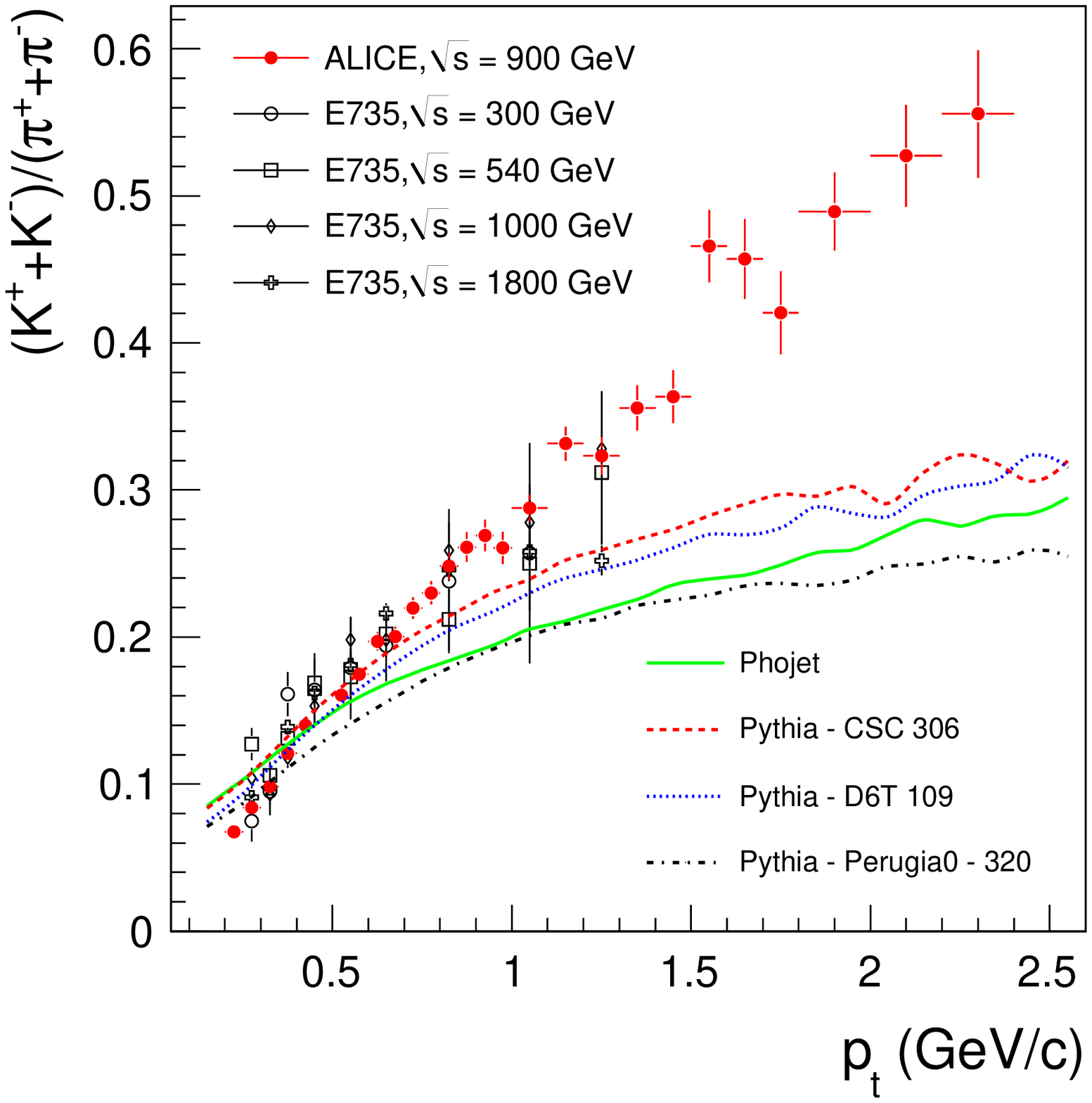}
\put(17,17){ALICE Preliminary}
\end{overpic}
\caption{Shown are kaon over pion ratios from ALICE and E735 compared to MC predictions.} \label{k_over_pi}
}

ALICE has measured inclusive transverse-momentum spectra of identified hadrons, $\pi^\pm$, $K^\pm$, and p($\pbar$), using three detector systems, the Inner Tracking System, the Time-Projecting Chamber and the Time Of Flight Detector. These systems cover different momentum regions and their results are consistent where these regions overlap. As examples, Figure~\ref{protons} shows the yield of protons and Figure~\ref{k_over_pi} the kaon over pion ratio. Both are compared to MC predictions which have difficulties describing the spectra. While the pion yield is reasonably described by PYTHIA D6T and Perugia-0 as well as PHOJET, kaons are underestimated up to a factor 2 for $\pt \gtrsim \unit[1]{GeV/\emph{c}}$ (plots not shown). Protons are also underestimated at higher $\pt$ except by PYTHIA D6T. The ratios, kaons over pions and protons over pions (plot not shown) are not described well qualitatively by the MCs considered.
It should be noted, however, that most of the MC tunes have not been extensively tuned to yields of identified particles.

The distributions can be fitted successfully with a Levi (Tsallis) function
which is of the form $dN_{\rm ch}/d\pt \propto \pt (1+(m_{\rm T}-m)/nT_l)^{-n}$ with the two parameters $n$ and $T_l$. Such a fit allows, for example, to extract the integrated yields.


\bfig
    \includegraphics[width=0.48\textwidth]{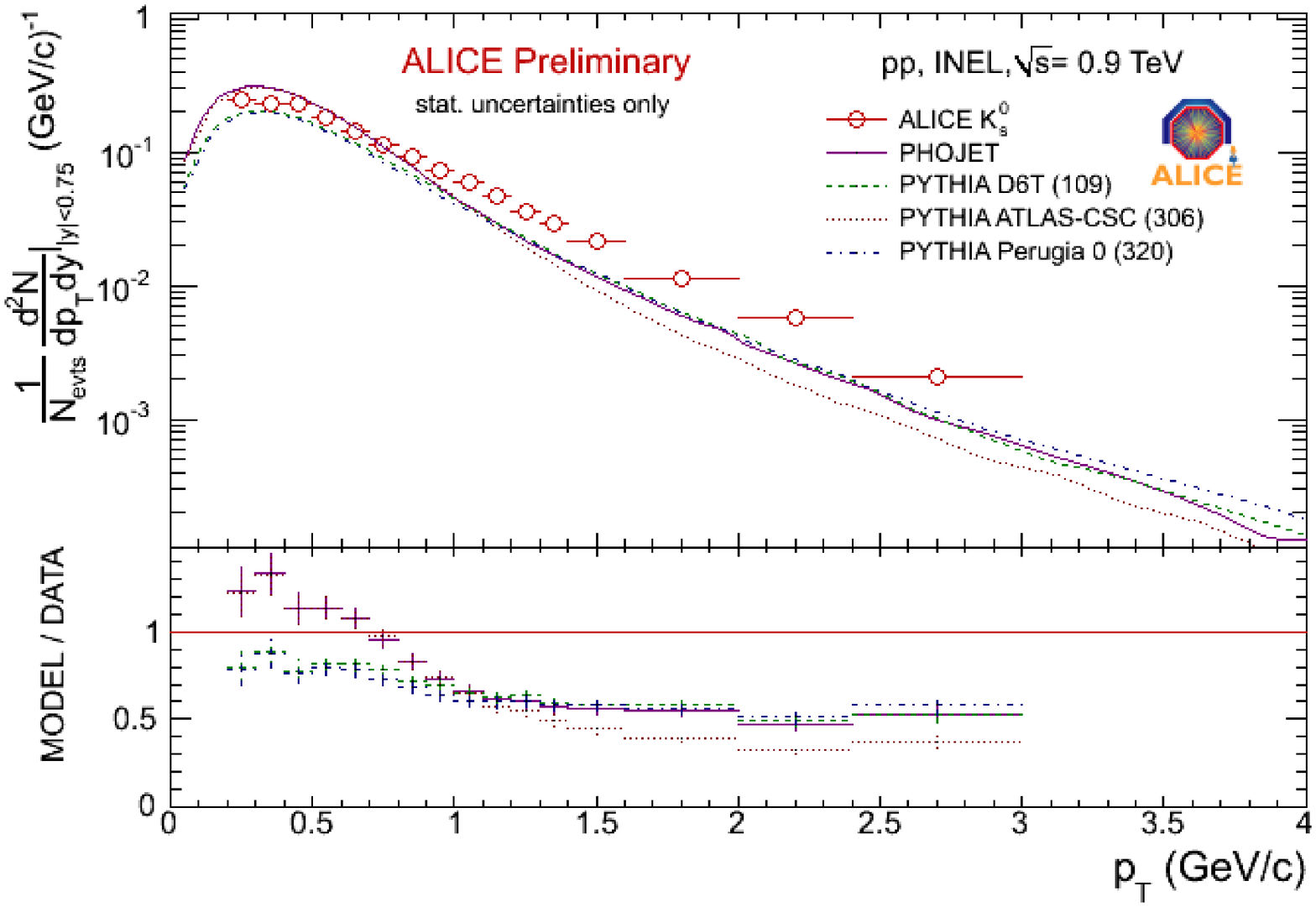}
    \hfill
    \includegraphics[width=0.48\textwidth]{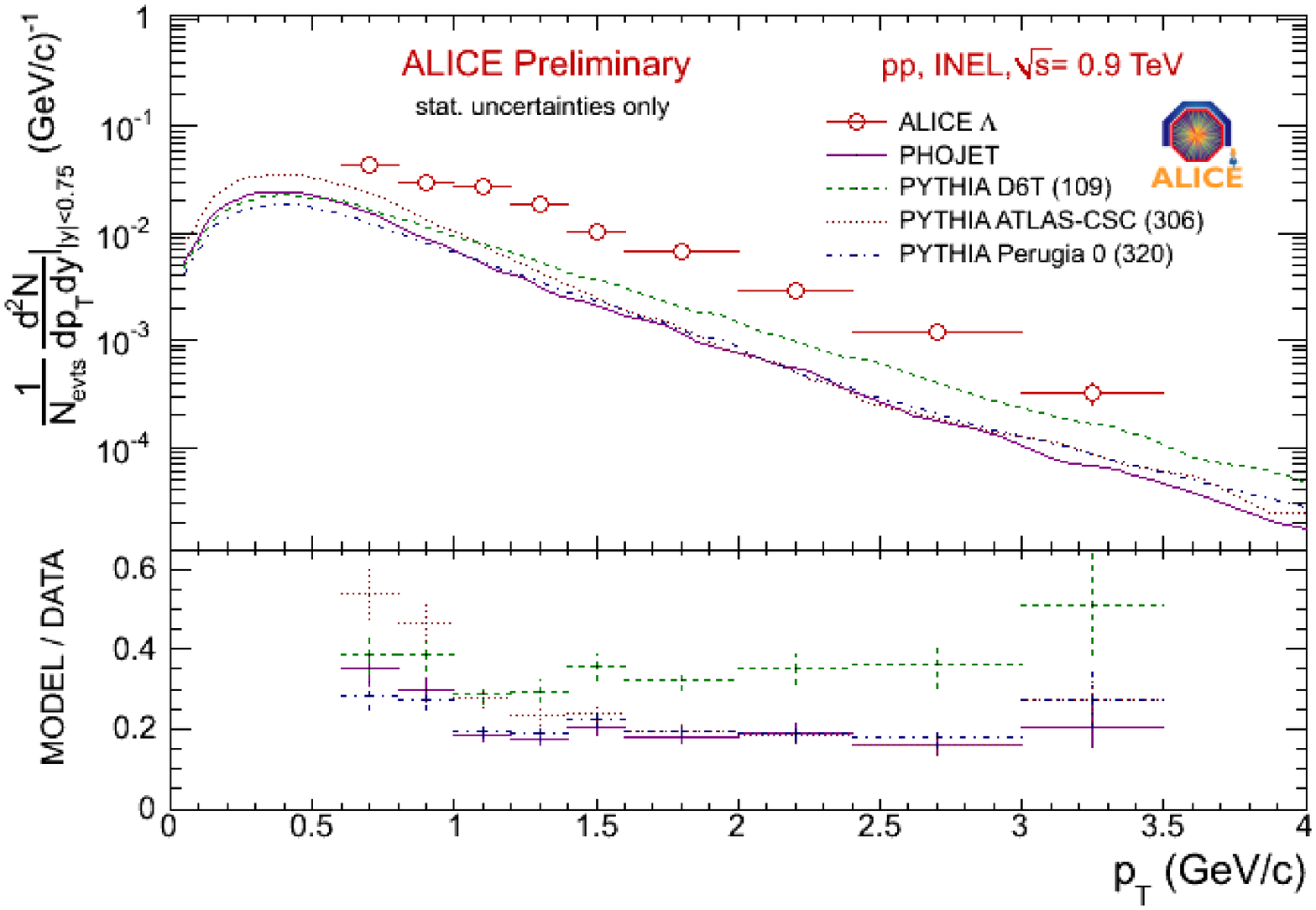}
    \caption{Shown are the yields of $K_0^S$ (left panel) and $\Lambda$ (right panel) compared to MC predictions. The lower panels shows the ratios MC over data. \label{strangeness} }
\efig

Yields of strange particles, i.e., $K_0^S$, $\Lambda$, $\Xi$, $\phi$ and their antiparticles, have been measured by combining the information of the PID detectors and evaluating the consistency of the potential daughter particle kinematics with the expected decay topology. Exemplarily, Figure~\ref{strangeness} shows the yields of $K_0^S$ and $\Lambda$ compared to MC predictions. Kaons are underestimated by about a factor 2 and $\Lambda$s up to a factor 5. Not shown is the comparison of $\Xi$ which is also about a factor 5 underestimated. Yields of strange particles are underestimated in general, however, the measured $\phi$ yield shows that it is described within uncertainties by the MC predictions (plot not shown). As previously, it should be noted that these MCs were not specifically tuned to reproduce measured strangeness yields.

\section{Summary \& Outlook}

ALICE has measured in the first year of LHC data-taking a comprehensive list of observables in minimum-bias pp collisions. Further measurements are ongoing including those that require higher statistics or special triggers. Examples are J/$\psi$ spectra and heavy flavor measurements, the measurement of the underlying event activity, jet and $\pi^0$ measurements. In addition ALICE started taking data with a high-multiplicity trigger that allows to assess the tail of the multiplicity distribution, in order to compare high-multiplicity events to bulk properties of minimum-bias events, like the average-$\pt$ and particle yields.

The measurements presented show that ALICE is operating well and ready for the start of the heavy-ion program which is expected at the end of 2010.

\end{document}